\documentclass[12pt]{article}
\usepackage{amssymb}
\usepackage{amsmath}
\usepackage{amsfonts}
\usepackage{graphicx}
\usepackage{mathrsfs}
\usepackage{comment}

\newcommand{\Slash}[1]{{\ooalign{\hfil#1\hfil\crcr\raise.167ex\hbox{/}}}}

\begin{document}
\begin{titlepage}
\begin{center}

\hfill TU-999\\

\vspace{2.5cm}

{\Large\bf 
Natural Split Mechanism for Sfermions: \\
$N$=2 Supersymmetry in Phenomenology}
\vspace{2.5cm}

 {\bf Yasuhiro Shimizu$^{(a)}$\footnote{email: yshimizu.hep@gmail.com}} and  {\bf  Wen Yin$^{(b)}$\footnote{email: yinwen@tuhep.phys.tohoku.ac.jp}}

\vspace{1.5cm}
{\it
$^{(a)}${\it  Academic Support Center, Kogakuin University, 

Hachioji  192-0015, Japan} \\
$^{(b)}${\it Department of Physics, Tohoku University, Sendai 980-8578, Japan} \\
}
\vspace{1.5cm}

\begin{abstract}
We suggest a natural split mechanism for sfermions based on $N$=2 supersymmetry (SUSY).
$N$=2 SUSY protects a sfermion in an $N$=2 multiplet from gaining weight by SUSY breaking.
 Therefore, if partly $N$=2 SUSY is effectively obtained, a split spectrum can be realized naturally. 
 As an example of the natural split mechanism, we build a gauge-mediated SUSY 
 breaking-like model assuming $N$=2 SUSY is partly broken in an underlying theory.  
The model explains the Higgs boson mass and muon anomalous magnetic dipole 
 moment within $1~\sigma$ level with a splitting sfermion spectrum. The model has seven light sparticles described by three free parameters and predicts a new chiral multiplet, sb: the $N$=2 partner of the $N$=1 ${\rm U(1)_Y}$ vector multiplet. The bini, the fermion component of the sb, weighs MeVs. We mention the experimental and cosmological aspects of the model. 
\end{abstract}

\end{center}
\end{titlepage}
\setcounter{footnote}{0}

\section{Introduction}

After the discovery of the Standard Model (SM) Higgs-like boson at the Large Hadron Collider (LHC) \cite{Aad:2012tfa},
no new particle has been found. On the other hand, some deviations from the SM have been observed in flavor physics \cite{Amhis:2014hma}.
If nature has supersymmery (SUSY), the current experiments suggest an interesting possibility: a split spectrum for sfermions. 
No discovery of sparticles \cite{dEnterriaCMS:2015jva}
 gives bounds to the squark masses: the squarks should be heavier than $\mathcal{O}(1)$ TeV. 
This is consistent with the simplest SUSY SM (minimal supersymmetric standard model: MSSM) 
prediction, since stops are required to be heavier than $\mathcal{O}$(10) TeV to explain the Higgs 
boson mass of about 125 GeV unless the SUSY breaking stop-Higgs trilinear coupling is 
parametrically large \cite{Okada:1990vk}.
In the case, the heavy sfermions automatically solve/alleviate the flavor/CP problem, the proton decay 
problem, several cosmological problems etc. 
On the other hand, if some deviations in flavor physics are true, there should also be light sfermions.
For example, the deviation of the muon anomalous magnetic dipole moment (muon $g-2$) from the SM prediction,
$\delta({g-2 \over 2})= (26.1\pm 8.0)\times 10^{-10}$ \cite{Bennett:2006fi,Hagiwara:2011af},
is above the $3~\sigma$ level. To relax the tension, a smuon as light as $\mathcal{O}$(300) GeV is needed in the MSSM \cite{Moroi:1995yh}.

A split spectrum commonly has the naturalness problem proposed by 't Hooft 
\cite{Hooft1979}.
If the particle mass is naturally small, by the definition of naturalness, there should be an approximate symmetry dominantly broken by the mass. 
The split spectrum mostly considered is the mass split between gauginos \cite{Ibe:2012hu} (gauginos with higgsino \cite{ArkaniHamed:2004yi}) and all the scalars except for the Higgs boson.
The split is natural since chiral symmetry forbids the mass term of a gaugino/higgsino. 
In SUSY SMs, the sfermion mass comes from SUSY breaking, hence the lighter sfermions should be protected from the SUSY breaking by a symmetry while the heavier ones are not\footnote{Note that the split between sfermions seems to be unconstrained by the anthropic principle. This is the difference from the case of the electroweak scale and the cosmological constant \cite{Agrawal:1997gf}.}.

In this paper, we suggest a natural split mechanism
using $N$=2 SUSY \cite{Grimm:1977xp}.
The non-renormalization theorem of $N$=2 SUSY forbids the sfermion mass terms which are embedded in $N$=2 multiplets even with SUSY slightly broken (Sec.\ref{chap:N2}).
If we add a sector with $N$=1 SUSY weakly coupled to the $N$=2 sector, the SUSY breaking dominantly affect the sfermion masses in the $N$=1 sector, while sub-dominantly to those in the $N$=2 sector since they are protected by $N$=2 SUSY. 
Therefore, a partly $N$=2 SUSY model can generate a naturally splitting sfermion spectrum.

 As an example, we construct a natural split gauge-mediated SUSY breaking (GMSB)  model \cite{Giudice:1998bp}
to explain the muon $g-2$ anomaly within the 1$\sigma$ level. 
We introduce the hyperpartners of left-handed sleptons, smuons, and down-type Higgs, with
 the extension of the ${\rm U(1)_Y}$ gauge interaction to $N$=2 SUSY by introducing a singlet chiral multiplet. 
 To cancel the gauge anomaly, we introduce spectator fields which are the mirror particles of the hyperpartners.  The messenger sectors are also partly extended to $N$=2 SUSY so that the natural split mechanism works. The condition of gauge coupling unification (GCU) at the SUSY GUT scale, $\sim 10^{16}$GeV, naturalness, and the symmetries require the additional charged fields to be at the same scale, the messenger scale. Below the messenger scale, the MSSM particles and the singlet survive.

In our model, there are only three free parameters: the messenger scale, the dominant and sub-dominant $F$-terms. The dominant one only gives masses to the sfermions in the $N$=1 sector at the leading 
order due to the natural split mechanism. The sub-dominant one acts as the one in the usual GMSB model.  We show an IR spectrum containing various testably light MSSM sparticles such as smuons, left-handed sleptons, and gauginos. 
In addition, there is a new light singlet fermion, bini, which is the ${\rm SU(2)_R}$ partner of the
 bino. 
In our model, the bini mass is $\mathcal{O}$(10) MeV, which is 2-loop suppressed to the bino mass.
The light bini rarely interacts with the MSSM particles, due to the higher dimensional interactions suppressed by the messenger scale.

\setcounter{footnote}{0}

\section{Natural Split Mechanism}

\label{chap:N2}

A sfermion mass term comes from the bilinear K{\" a}hler term,  
\begin{align}
\label{eq:Kmass}
\mathscr{K}={|Z|^2 \over M^2}Q^{\dagger}Q \supset |\theta|^4 \left|{F_Z \over M}\right|^2   \tilde{Q}^{\dagger} \tilde{Q}.
\end{align}
We will show $N$=2 SUSY forbids such a term. 

An $N$=2 SUSY extension of the SM is hardly considered for phenomenology. 
There are at least two difficulties.
The renormalizable $N$=2 SUSY theory only has fermions that are in real representations, while the SM fermions are chiral. The Landau-pole of the gauge coupling is generated at a low energy scale. 
However, a partly $N$=2 SUSY extension at a high energy scale is not so problematic.

\subsection{Review of $N$=2 SUSY}

Here we briefly review $N$=2 SUSY.
For simplicity, we consider $N$=2 SUSY quantum electrodynamics.

\label{chap:N2SUSY}

The Lagrangian is given in the $N$=1 superfield formalism as
\begin{align}
\label{eq:N2lagabe}
\begin{split}
\mathscr{L}_{N=2} &= \mathscr{L}_{matter}+\mathscr{L}_{gauge},\\
\mathscr{L}_{matter} &= \int{d^4\theta\sum_{i}^{N_f} (\overline{\Phi}_i e^{-2Y_i V}\overline{\Phi }^\dagger_i+\Phi^\dagger_i e^{2Y_i V}\Phi_i)}\\
&+ \int{d^2\theta \sum_{i}^{N_f}( \sqrt{2}\overline{Y}_i \overline{\Phi}_i\Phi_Y\Phi _i+\sum_j{m_{i,j} \overline{\Phi}_i \Phi _j})}+{\rm h.c.},\\
\mathscr{L}_{gauge} &= ({1 \over g^2}\int{d^4\theta {\Phi_Y^\dagger \Phi _Y}}+{1 \over 4g^2}\int{d^2\theta W_{\alpha}W^{\alpha}}+{\rm h.c.}),\\
\end{split}
\end{align}
with $ Y_i =|\overline{Y}_i|,\ [m^{\dagger},m]=0$.
The Lagrangian is the most general non-trivial renormalizable Lagrangian with $N$=2 SUSY without the Fayet-Iliopoulos (FI) term \cite{AlvarezGaume:1996mv}.

There are two kinds of $N$=2 multiplets in the Lagrangian: vector multiplet and hypermultiplet. The Lagrangian has a specific symmetry, ${\rm SU(2)_R}$ symmetry, that rotates the fields in the $N$=2 multiplet as in Tab.\ref{tab:multiplet}.
\begin{table}[h]
\begin{center}
\begin{tabular}{|c|c|c|c|}\hline $N$=2 Multiplet & in Superfields & in Fields & renormalized at \\\hline {\bf Vector multiplet}  &($V, \Phi_Y$) & $\left((A^\mu, \lambda,D),( \Phi_Y,\psi_Y,F_Y)\right)$ & 1-loop level \\\hline {\bf Hypermultiplet}& ($\Phi_i,\overline{\Phi}_i^\dagger $) &  $((\phi_i,\psi_i,F_i ), (\overline{\phi}_i, \overline{\psi}_i,\overline{F}_i)^\dagger) $ & tree level
 \\\hline \end{tabular} 
 \begin{tabular}{|c|c|c|c|c|c|}\hline ${\rm SU(2)_R}$ Multiplet & ($\phi_i,\overline{\phi}_i^{\dagger}$) & ($\lambda, \psi_Y$) &($\overline{F}_i^\dagger,F_i$) &$(\Re{[F_Y]},\Im{[F_Y]}, {1\over \sqrt{2}}D$)&the others \\\hline Representation &{\bf 2} & {\bf 2}& {\bf 2}&{\bf 3}&1
 \\\hline \end{tabular} \end{center} 
 \caption{$N$=2 multiplets and ${\rm SU(2)_R}$ representations}
\label{tab:multiplet}
 \end{table}
The ${\rm SU(2)_R}$ symmetry is not only non-commutative with $\theta$, but also not closed in an $N$=1 supermultiplet.

Since the Lagrangian has the ${\rm SU(2)_R}$ symmetry only if $Y_i=|\overline{Y}_i|$,
 the ${\rm SU(2)_R}$ symmetric models do not allow $\overline{Y_i}$ to vary even with the
wave functions renormalized.
Because of the non-renormalization theorem for the superpotential, the ${\rm SU(2)_R}$ symmetry leads to the non-renormalization theorem to the K{\" a}hler potential as in the Tab.\ref{tab:multiplet}.

\subsection{$N$=2 SUSY breaking}
\label{chap:N2SB}
We consider the case that $N$=2 SUSY is slightly broken to $N$=0 by an $F$-term, $F_Z$, of a SUSY breaking field, $Z$.
Since $N$=2 SUSY is recovered with the vanishing $F$-term, the effective Lagrangian should be
\begin{align}
\begin{split}
\mathscr{L}_{\rm eff}&=\mathscr{L}_{N=2}+\mathscr{L}_{N2B},\\
\mathscr{L}_{N2B}&=\int{d^4\theta \tilde{W}(\Phi_i,\overline{\Phi}_i,\Phi_Y,Z,Z^\dagger) +{\rm h.c}} \ .
\end{split}
\end{align}
Here $\tilde{W}$ is a holomorphic function of $\Phi_i,\overline{\Phi}_i$, and $\Phi_Y$. 
Therefore, even with SUSY breaking, sfermion masses like Eq.(\ref{eq:Kmass}) are forbidden by the non-renormalization theorem.
On the other hand, integrating $\overline{\theta}^2$ out, an effective superpotential, $F_{Z}^\dagger\partial_{Z^\dagger}\tilde{W}$, is generated.

If $\Phi_Y$ is a gauge singlet as in Sec.\ref{chap:N2SUSY}, the dangerous tadpole term, $\delta\tilde{W} \sim Z^\dagger \Phi_Y$, may be generated.
$Z^\dagger \Phi_Y$ can be a source of large $\Phi_Y \sim Z$ or $F_Y \sim  F_Z$ unless fine-tunings.
To have $<\Phi_Y> \sim 0$ naturally, there are two possibilities. 
One is to introduce a large SUSY preserving mass term, $W=M\Phi_Y^2$, by hand to suppress $<\Phi_Y>$.
The other is to impose a symmetry to forbid the tadpole term or to have a naturally small one.
In this paper, to forbid the tadpole term we assume a symmetry, under which $Z$ is charged, while $\Phi_Y$ is not.

A concrete $Z_2$ symmetric superpotential with slightly broken $N$=2 SUSY is as follows.
\begin{align}
\label{eq:SBz2}
W&=\sqrt{2}Y\Phi_Y(\Phi_1 \overline{\Phi}_{1}+\Phi_{-1} \overline{\Phi}_{-1})+Z( \Phi_1 \overline{\Phi}_{1}- \Phi_{-1} \overline{\Phi}_{-1}).
\end{align}
The model possess a $Z_2$ symmetry which exchanges the indices and reverses the sign of $Z$.
Since $Z$ can be identified as the chiral component of an $N$=2 vector multiplet\footnote{Actually, non-linear $N$=2 abelian gauge theory with the electric and magnetic FI terms aligned in the ${\rm SU(2)_R}$ basis, can spontaneously break $N$=2 SUSY to $N$=0 with the chiral component of the $N$=2 vector multiplet acquiring SUSY breaking VEV, $Z$ \cite{Antoniadis:1995vb}.}
 $F_Z$ breaks the ${\rm SU(2)_R}$ to ${\rm SO(2)}$, which also forbids the sfermion mass terms, Eq.(\ref{eq:Kmass}). 

\subsection{Natural split mechanism} 
\label{chap:mech}
We explain the essence of the natural split mechanism.
Suppose we have three sectors\footnote{A sector means that it is separated if some weak couplings are taken to zero.}: SB sector, L sector and H sector. 
SUSY is broken in the SB sector, while the other sectors do not directly couple with the SUSY breaking.
At the SUSY limit the SB and L sectors have $N$=2 SUSY, while the H sector has $N$=1 SUSY.

\begin{figure}[!h]
\begin{center}
   \includegraphics[width=100mm]{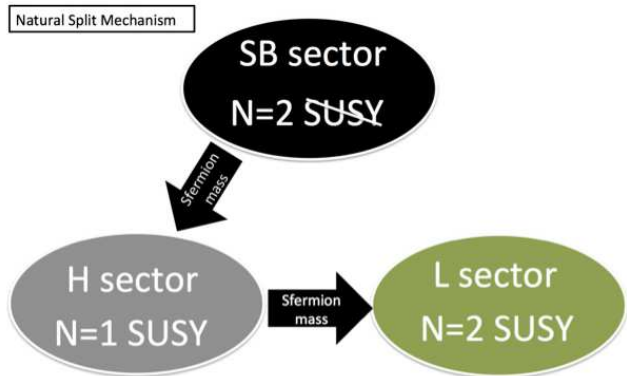}
\caption{The sketch of the natural split mechanism}
\label{fig:mech}
\end{center}
\end{figure}

To have the SUSY breaking in the L sector, the SUSY breaking effect should be mediated firstly to 
the H 
sector to pick up the explicit $N$=2 SUSY breaking, and then be mediated to the L sector.
The sfermion masses in the L sector are suppressed to those in the H sector. 
Therefore, integrating out the SB sector, we obtain a split spectrum for sfermions.
If the explicit breaking of $N$=2 SUSY is small enough, the ${\rm SU(2)_R}$ partners of the 
light sparticles are also light, which may imply new light particles.
The natural split mechanism is summarized in Fig.\ref{fig:mech}.

The set-up is natural, if the L 
and SB sectors have small deviations from the SO(2) relations, in the sense of Eq.(\ref{eq:SBz2}). Then, there are two explicit breaking sources of the SO(2): the deviations and the 
existence of the H sector. It is natural (or stable under the quantum correction), when the two sources are the same order. For example, if the H sector particles contribute to the reactions in the L and SB sectors via loop effects as the case we will consider,  
it is still natural to have the deviations of the SO(2) in the L and SB sectors as small as the loop effects. In the case, the 
split spectrum for sfermions is not spoiled since the sfermion masses from the SO(2) deviations in the L and
SB sectors is also suppressed.

The assumption of partly $N$=2 SUSY seems to be reasonable. 
This is because a series of SUSY theories have the possibility of partly $N$=2 SUSY as an effective 
theory. It may be obtained explicitly or spontaneously. 

The compactified extra-dimensional theory with ``our world" localized on 
an orbifold can be the former case. For example, $N$=1 SUSY on $R_{1,3} \times S_2/Z_2$  
spacetime is effectively $N$=2 SUSY which is broken by $Z_2$ projection \cite{ArkaniHamed:2001tb}.  If ``our world" is localized on the singularity, only $Z_2$ even fields survive at the low energy limit. 
Ordinary, to get $N$=1 SUSY, the $N$=2 partner of the MSSM particles are assumed $Z_2$ odd 
\cite{ArkaniHamed:2004yi, Schmaltz:2000gy}. However, our world may contain a part of a completely $Z_2$ even $N$=2 multiplet whose partner also survives at the limit. 
Therefore, partly approximate $N$=2 SUSY is obtained.

We may also break $N$=2 SUSY to $N$=1 spontaneously since $N$=1 gauge theory can be 
described by non-linear realized $N$=2 SUSY with chiral matters of any representations \cite{Bagger:1994vj}. In \cite{Antoniadis:1995vb}, an $N$=2 non-linear abelian gauge model is shown to have such 
breaking with electric and magnetic $N$=2 FI terms in a generic case. Therefore, a separated 
sector, which does not directly couple to the $N$=2 gauge fields, has $N$=2 breaking only at the higher 
order.

Although the consistency of the above possibilities need to be investigated concretely in our future work,
we assume a situation that $N$=2 SUSY is broken partly by some underlying physics in this paper. 

\setcounter{footnote}{0}
\section{Natural Split GMSB Model}
\label{chap:p5}
In this section, we show a concrete model with the natural split mechanism.

We assume the effective MSSM, GCU at the SUSY GUT scale  $\sim 10^{16}$GeV,
and the explanation of the muon $g-2$ anomaly within the 1$\sigma$ level \cite{Bennett:2006fi, Hagiwara:2011af,Moroi:1995yh}. 
A GMSB model is a natural candidate to satisfy the assumptions, however as noted in \cite{Ibe:2012qu}, a split spectrum is difficult to obtain due to the non-zero hypercharge of the messengers except for those belonging to the adjoint representation of SU(5). The natural split mechanism can generate a split spectrum even with only hypercharged messengers.

One of the advantages of the GMSB mechanism in our case is that
the ordinary messenger fields are able to be the hyperpartners of the muon superfields. 
The other advantage is as follows.
To keep light smuons natural in the effective MSSM, the following relations are required from the 1-loop renormalization group equations (RGEs) \cite{Martin:1993zk}. 
\begin{align}
\label{eq:natura}
m_{\rm smuon}^2 &\gtrsim ({Yg'^2}S, {Y^2g'^2}M_1^2,{g^2}M_2^2),\\
{\rm where}\
\label{eq:S}
S&=\sum_i^{\rm DOF}{m_i^2 Y_i}=\rm{Tr}{[m_{\tilde{d}}^2+m_{\tilde{e}}^2-m_{\tilde{L}}^2+m_{\tilde{Q}}^2-2 m_{\tilde{u}}^2]-m_{H_d}^2+m_{H_u}^2}.
\end{align}
Since the vanishing of the right hand side of Eq.(\ref{eq:natura}) implies that smuons are separated from the other sectors, the zero limit of the smuon masses is the recovery of SUSY in the separated smuon sector. Therefore, we should have a light wino, a light bino and a small $S$-term. To obtain a small $S$-term, there should be
a large cancellation of the heavy sfermion masses including the stop masses $\gtrsim$$\mathcal{O}$(10) TeV. In the GMSB mechanism, the $S$-term is guaranteed to vanish automatically \cite{Meade:2008wd}. The gauginos are naturally light if the messenger sector has an approximate discrete R 
symmetry.

\subsection{The set-up of the natural split GMSB model}
The Lagrangian density just above the messenger scale, $M$, is as follows.
\label{chap:Model}
\begin{align}
\label{eq:L}
\begin{split}
\nonumber
\mathscr{L}=&\int{d^4\theta \left(K_{\rm MSSM}+K_{\rm N1Mess}+K_{\rm N2Mess}+K_{\rm Partner}+K_{\rm Spectator}\right)}\\
&+\int{d^2\theta \left(W_{\rm MSSM}+W_{\rm N1Mess}+W_{\rm N2Mess}+W_{\rm N2}+W_{\rm mass}\right)}
\end{split}\\
&+{\rm h.c.} +\mathscr{L}_{\rm gauge},\\
K_{\rm MSSM}&=\sum_{i=1}^{3}{(|Q_i|^2+|u_i|^2+|d_i|^2+|L_i|^2+|e_i|^2)}+|H_u|^2+|H_d|^2, \\
K_{\rm N1Mess}&=\left( |\Phi_u|^2 + |\overline{\Phi}_u|^2 +|\Phi_Q |^2+|\overline{\Phi}_Q|^2\right),\\
K_{\rm N2Mess}&= \sum_a^{+,-}\sum_{i}^{-1,1}{(|\Phi_{di}^{a}|^2+|\overline{\Phi}_{di}^{a}}|^2),\\
K_{\rm Partner}&={1 \over g'^2} |\Phi _Y|^2 +|\overline{e_{2}}|^2+|\overline{H_{d}}|^2+\sum_{i=1}^3{| \overline{L_{i}}|^2},\\
K_{\rm Spectator}&=|S_{e2}|^2+|S_{Hd}|^2+\sum_{i=1}^3{| S_{Li}|^2}, \\
W_{\rm MSSM}&\sim y_{t}H_{u}Q_{3}u_{3}+y_{b}H_{d}Q_{3}d_3+y_{\tau}H_{d}L_{3}e_3,\\
W_{\rm N1Mess}&= \lambda_u Z\Phi_u \overline{\Phi}_u +\lambda_Q Z\Phi_Q \overline{\Phi}_Q,\\
W_{\rm N2Mess}&=\sum_a^{+,-}{\left( \sum_i^{-1,1}{({1\over \sqrt{2}}{1\over 3}\Phi_Y \Phi_{di}^a\overline{\Phi}_{di}^a})+Z_a(\Phi_{d1}^a\overline{\Phi}_{d1}^a-\Phi_{d(-1)}^a\overline{\Phi}_{d(-1)}^a)\right)},
\end{align}
\begin{align}
W_{\rm N2} &=-{1\over \sqrt{2}}\int{d^2\theta \Phi_Y \left({1 \over 2} \sum_{i=1}^3{ L_i \overline{L}_i}+{1\over 2}H_d\overline{H_d}- e_2 \overline{e}_2 \right)}+{\rm h.c},\\
W_{\rm mass} &=M\left( \sum_i^3{\overline{L_i}S_{{Li}}}+\overline{H}_d S_{Hd}+\overline{e}_2 S_{e2} \right),
\end{align}
where
\begin{align}
\begin{split}
Z&=M+\theta^2 F=M(1+\theta^2\Lambda ), \\
 Z_+&=M+\theta^2 F_d=M(1+\theta^2\Lambda_d ),\\ 
 Z_- &=M-\theta^2 F_d=M(1-\theta^2\Lambda_d ),
\end{split}
\end{align}
with $\Lambda_d \gg \Lambda$ and $\lambda_{u,Q}\sim \mathcal{O}(1)$. Here, we omit to 
write down the terms with $N$=1 vector multiplets. 
In the model, the ${\rm U(1)_Y}$ gauge interaction are partly extended to $N$=2 SUSY with 
introducing the $N$=2 partners: $\overline{L_i}$, $\overline{e_2}$, $\overline{H_d}$ and $\Phi_Y$. The 
spectator fields, $S_{Li},S_{e2}$ and $ S_{Hd}$, are needed to cancel the gauge 
anomalies and to obtain the mass terms of the introduced hyperpartners to satisfy the assumption of 
effectively MSSM.
There are two messenger sectors: the ordinary one and the one with the ${\rm U(1)_Y}$ gauge couplings 
extended to $N$=2. The latter one couples with SUSY breaking fields, $Z_+$ and $Z_-$, with the $F$-term much larger than that of the former one, $Z$. The doubling of $Z_+$ and $Z_-$ are needed in 
order to cancel the large gaugino masses by an approximate $Z_{4R}$ symmetry\footnote{Such an $N$=2 $\rightarrow$ $N$=0 SUSY breaking model can be realized if we have two $Z_2$ related $N$=2 abelian vector multiplet systems in \cite{Antoniadis:1995vb} with the same prepotentials but opposite FI terms. 
Since the $N$=2 $D$-term is proportional to the FI term and the scalar potential does not care about the sign of the FI term, the two $F$-terms of the chiral components are opposite while the scalar VEVs, if no degeneracy, are the same. If $\Phi_{di}^+$ and $\Phi_{di}^-$ couple with the two $N$=2 vector multiplets respectively, the 
 $Z_{+}$ and $Z_{-}$ can be recognized as the VEVs of the chiral components of them.}.
To forbid the tad-pole term of the $N$=2 ${\rm U(1)_Y}$ vector partner, $\Phi_Y$, 
namely sb, a precise $Z_2$ symmetry is also needed. The components and their properties are shown in Tab.\ref{tab:sym}.

\begin{table}[!h]
\begin{center}
\begin{tabular}{|c|c|c|c|c|c||c|c|c|}\hline \multicolumn{1}{|c|}{$N$=2 sector}& \multicolumn{4}{|c|}{ MSSM particle and $N$=2 partner}& \multicolumn{1}{|c||}{Messenger}&\multicolumn{3}{|c|}{Spectator}
\\\hline Component & $V_Y$,$\Phi_Y$ & \multicolumn{1}{|c|}{$L_i$,$\overline{L_i}^\dagger$}&\multicolumn{1}{|c|}{$H_d$,$\overline{H_d}^\dagger$}&$e_2$,$\overline{e}_2^\dagger$ &$\Phi_{di}^a$,$\overline{\Phi}_{di}^{a\dagger}$ & $S_{Li}$&$S_{Hd}$ &$S_{e2}$  \\\hline 
SM Gauge&$[{\bf 1},{\bf 1},0]$&\multicolumn{2}{|c|}{[{\bf 1},{\bf 2},-${1 \over 2}]$}&[{\bf 1},{\bf 1},1]&[$\overline{{\bf3}}$,{\bf 1},${1 \over 3}$]&\multicolumn{2}{|c|}{[{\bf 1},{\bf 2},-${1 \over 2}$]}&[{\bf 1},{\bf 1},1]\\ \hline 
DOF & / & \multicolumn{1}{|c|}{3$\times$2}& \multicolumn{1}{|c|}{1$\times$2} & 1$\times$2 & 2$\times2\times$2 &3&1&1 \\ \hline 
\end{tabular}
\begin{tabular}{|c|c|c|c|c|c|c|c|c|c|c|}\hline \multicolumn{1}{|c|}{$N$=1 sector}& \multicolumn{5}{|c|}{ MSSM chiral field}& \multicolumn{4}{|c|}{Messenger}
\\\hline Component & $Q_i$ &$u_i$  &$d_i$ &$ e_1, e_3$&$H_u$ &$\Phi_{Q}$ &$\overline{\Phi}_{Q}^\dagger$&$\Phi_{u}$ & $\overline{\Phi}_{u}^\dagger$\\\hline 
SM Gauge&$[{\bf 3},{\bf 2},{1 \over 6}]$&$[\overline{\bf 3},{\bf 1},$-${2 \over 3}]$&$[\overline{\bf 3},{\bf 1},{1 \over 3}]$&$[{\bf 1},{\bf 1},{1 }]$&$[{\bf 1},{\bf 2},{1 \over 2}]$ & \multicolumn{2}{|c|}{$[{\bf 3},{\bf 2}, {1 \over 6}]$}& \multicolumn{2}{|c|}{$[\overline{\bf 3},{\bf 1},$-${2 \over 3}]$}  \\ \hline 
DOF & 3 & 3 & 3& 2 &1&1&1&1&1  \\\hline 
\end{tabular}
\begin{tabular}{|c|c|c|c|c|c|c|c|c|c|}\hline 
\multicolumn{8}{|c|}{ Action under Descrite Symmetry}\\\hline 
Field & $\Phi_Y$ & $Z_{\pm}$ & $Z$ &($\Phi_{di}^a$,$\overline{\Phi}_{di}^a$)&$\overline{\Phi}_Q$&$\overline{\Phi}_u$ & Others \\\hline 
$Z_2$& $\Phi_Y$ &$-Z_{\pm}$ &$-Z$& ($\Phi_{d(-i)}^a$,$\overline{\Phi}_{d(-i)}^a$) & $-\overline{\Phi}_Q$ & $-\overline{\Phi}_u$ & Itself  \\\hline
$Z_{4R}$& $\Phi_Y(i\theta)$ & $Z_{\mp}(i\theta)$=$Z_{\pm}(\theta)$ & $Z(i\theta)$ &i($\Phi_{di}^{(-a)}$,$\overline{\Phi}_{di}^{(-a)}$) & i$\overline{\Phi}_Q$ & i$\overline{\Phi}_u$ & =$\sqrt{Z_R}$ \\\hline
 \end{tabular}
  \caption{Particles and symmetries in the $N$=1 language. $Z_R$ is $R$-parity.}
 \label{tab:sym}
 \end{center}
\end{table}

We explain the reasons for the choice of the hyperpartners and the SUSY mass terms. 
The choice of $N$=2 sector is not artificial, but is the requirement of naturalness, which urge us to have 
a small $S$-term. 
The would-be large smuon masses mediated by the ${\rm U(1)_Y}$ should be almost 
cancelled by $N$=2 non-renormalization theorem as in Sec.\ref{chap:N2SB}. The GMSB effect via the $V_Y$ to a scalar mass is $\delta m_i^2 \propto Y_i^2$, hence the $S$-term is generated from the zero value of the GMSB mechanism: $
\delta S \propto -\sum_i{Y_i^3} =-2Y_{L_2}^3-Y_{e_2}^3=- {3 \over 4}$. To cancel the $S$-term, one of the solution is to have three additional cancellations of the $({\bf 1},{\bf 2},-{1 \over 2})$-type scalar masses. 
Therefore, we can have the hyperpartners as in Tab.\ref{tab:sym} at least. Suppose that the hyperpertners have Dirac mass terms with the spectators. Then the hyperpertners should decouple almost at the same scale, otherwise the cancellation of the $S$-term is spoiled by the RGEs.  
Therefore, three SUSY masses, $M$, $Z$ and $Z_{\pm}$, should be the same order\footnote{Since $M \lesssim Z_{\pm}$, we can show that in general, the GCU is 
 obtained at a scale $\lesssim 10^{16}$GeV. We have taken a most favored value by the proton decay 
 problem.}, when we assume the GCU at the SUSY GUT scale, since the hyperpartners, the messengers and the spectators can be embedded into complete SU(5) 
  multiplets, $({\bf 5+\bar{5}}) \times 4$ and ${\bf 10+\overline{10}}$. Note that as usual GMSB, the sfermion mass squared depends on $\Lambda^2,\Lambda^2_{d}$ linearly but the SUSY masses logarithmically accompanied by a loop factor \cite{Giudice:1998bp}. Therefore, the change of the relative factor $\sim 3$ between the SUSY masses, is sensitive only up to the next loop order to the sfermion masses. This allows us to set the SUSY masses, $M$, $Z$ and $Z_{\pm}$, to be the same value, $M$, without changing the result at the leading order\footnote{The bini mass actually has a power-law dependence on the relative factor, but in this paper we do not care about its value but the order.}. There is a bound for $M$ to be larger than $\mathcal{O}(10^6)$GeV so that the GCU remains perturbative.

\subsection{Natural split mechanism and low energy spectrum}
If we forget about the spectators and the $N$=1 messengers, the model corresponds to a complicated case of Fig.\ref{fig:mech}.
We can identify that the $N$=2 messenger sector is the SB sector, the hypermultiplets in the $N$=2 sector compose
the L sector, the $N$=1 and $N$=2 gauge multiplets are mediators, and the other fields compose the H Sector. 

\begin{figure}[!h]
 \begin{center}
\begin{minipage}{0.45\hsize}
 \includegraphics[width=75mm]{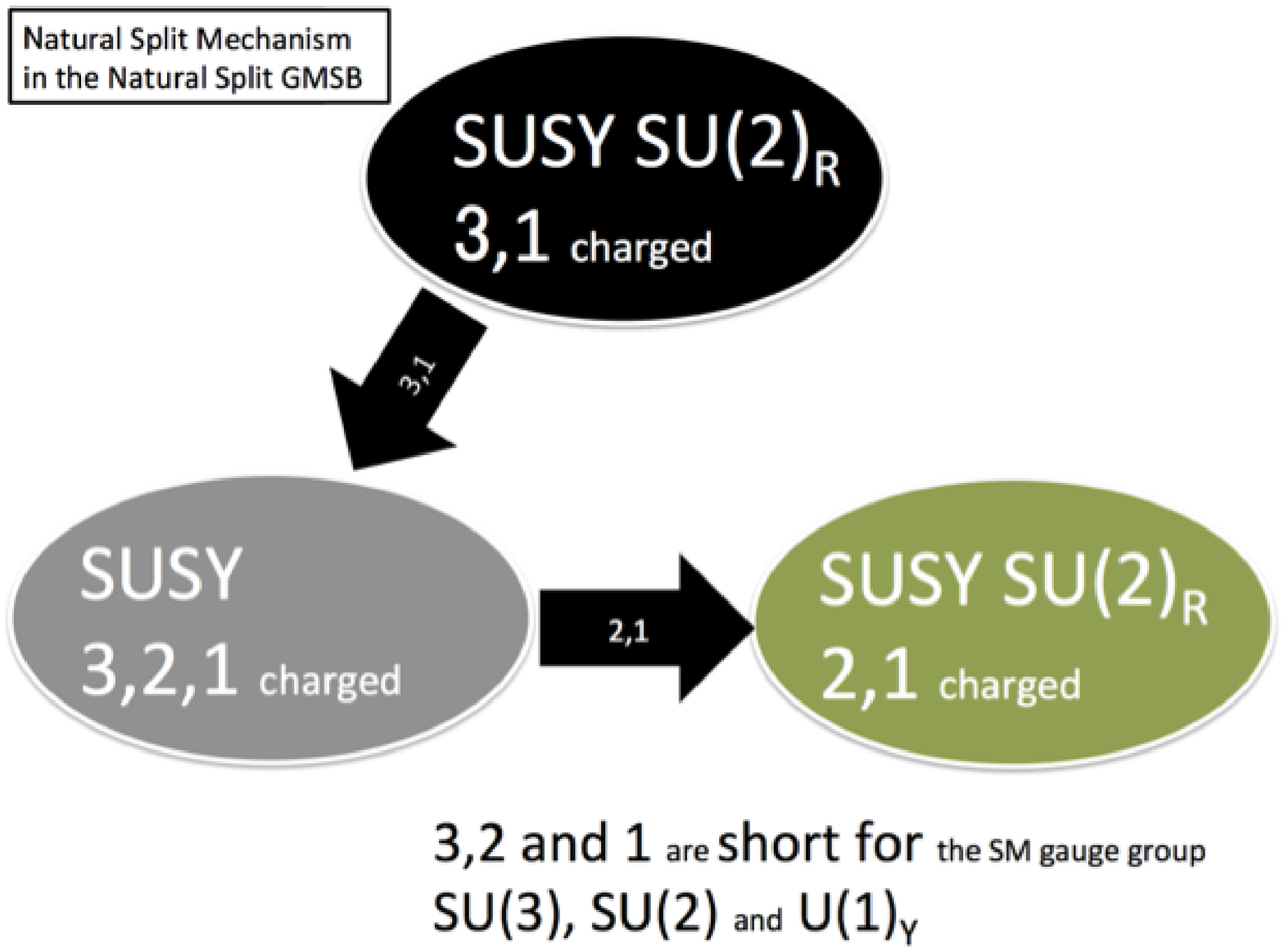}
\end{minipage}
\begin{minipage}{0.45\hsize}
   \includegraphics[width=75mm]{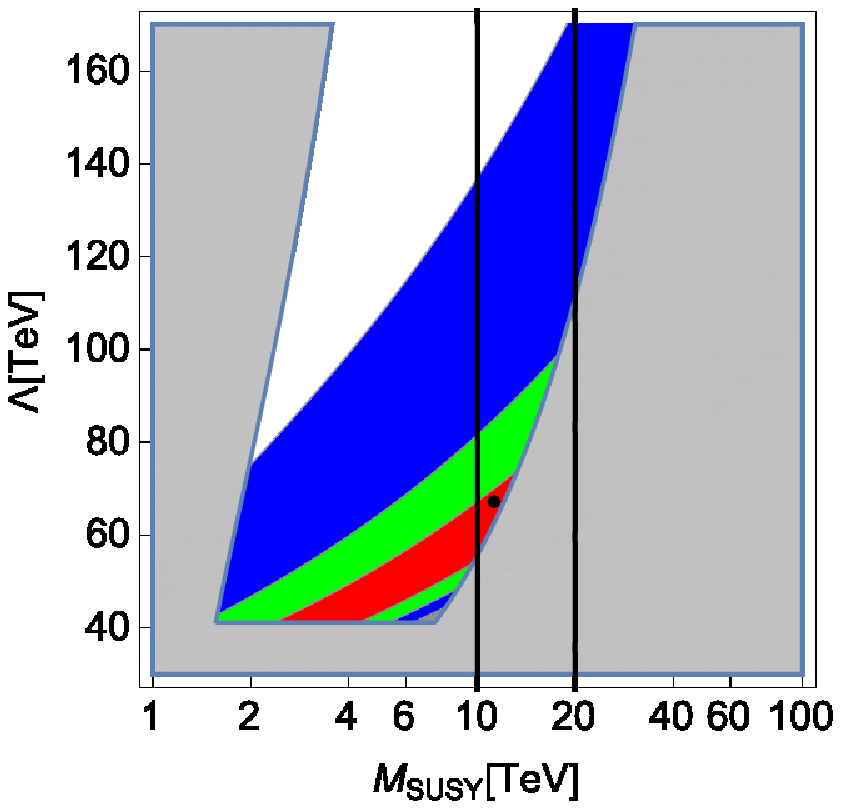}
\end{minipage}
  \caption{{\bf Left} The sketch of the mediation of the dominant SUSY breaking, $F_{d}$. The black (grey, green) ellipse is the SB (H, L) sector.
 {\bf Right} The parameter region of the natural split GMSB model with tan$\beta$=7, $M=10^{10}$ GeV. Between the two vertical black-lines, the Higgs boson mass is between 124.5 GeV and 127.2 GeV calculated by FeynHiggs 2.11.2 \cite{Hahn:2013ria} with about 1 GeV uncertainty. $M_{\rm SUSY}$ is the geometric average of the stop mass.
The red (green, blue) region explains the muon $g-2$ anomaly within the $1~\sigma$ ($2~\sigma$, $3~\sigma$) level. The black dot has the information in Tab.\ref{tab:Fpara}.}
\label{fig:GMSB}
\end{center}
\end{figure}

The natural hierarchy is obtained as in Tab.\ref{tab:UVspectra} by integrating out the degrees of freedom above 
the messenger scale.
 The gauge charges and the natural split mechanism forbid SUSY breaking to mediate into the L sector at the 
 leading order. The sfermion masses in the L sector are naturally suppressed to those in the H sector which is 
 mediated directly from the SB sector. The gauginos are also naturally light since the $Z_{4R}$ parity
forbids the gaugino masses, as well as the naturalness of small $F$ compared with $F_{d}$.

\begin{table}[!h]
\begin{center}\begin{tabular}{|c|c|c|c||c|c|c|c|c|c|}\hline  Sector  &  \multicolumn{3}{|c||}{ H Sector } & \multicolumn{5}{|c|}{Mediator Sector}    \\\hline Particle  & Squarks  & ${\rm \tilde{e}_1,\tilde{e}_3}$ &${\rm H_u}$ & Sb & Bini  & Bino  & Wino  & Gluino  \\\hline   Mass ${g'^2 \over 16 \pi^2}$$\cdot$ & $\sqrt{20 \over 3}{g_3^2 \over g'^2}\Lambda_d$ & $\sqrt{10 \over 3}\Lambda_d $   &  $\sqrt{5 \over 6} \Lambda_d$ &  ${ \sqrt{32} \over 3}{g_3 \over g' }\Lambda_d$ &${L^2}$$\Lambda$&${3 \over 2}\Lambda$ & ${3 \over 2}{g^2 \over g'^2}\Lambda  $&${3 \over 2}{g_3^2 \over g'^2 }\Lambda$   \\\hline \hline Sector  &  \multicolumn{8}{|c|}{ L Sector }    \\\hline Particle  &  \multicolumn{4}{|c|}{ ${\rm \tilde{L}_1},{\rm \tilde{L}_2},{\rm \tilde{L}_3},{\rm H_d}$} & \multicolumn{4}{|c|}{$\tilde{e}_2$}  \\\hline   ${\rm Mass^2}$ & \multicolumn{4}{|c|}{  $3({g^2 \over 16 \pi^2})^2 \Lambda^2$+${3 \over 4}({g'^2 \over 16 \pi^2})^2 \Lambda^2$+$L m_{{\rm H_u}}^2$} &  \multicolumn{4}{|c|}{$ 3({g'^2 \over 16 \pi^2})^2 \Lambda^2$+$L m^2_{{\tilde {\rm e}_{1,3}}}$} \\\hline \end{tabular} \caption{The formula of mass spectrum just below the messenger scale $M$ with assuming $\Lambda_d \gg \Lambda$ and neglecting the $\Lambda$ compared with $\Lambda_d$ \cite{Giudice:1998bp}. $L$ denotes the loop factor $\sim \mathcal{O}(10^{-2})$ which can be determined by the higher loop calculations.}
\label{tab:UVspectra}
\end{center}
\end{table}

With solving the RGEs of the MSSM, we show the allowed region in Fig.\ref{fig:GMSB} using the method 
developed in \cite{RECC}.  The black dot in the allowed region (Fig.\ref{fig:GMSB}) has the IR spectrum in Tab.\ref{tab:IRspectrum}. In our analysis, we treat $B_0$ as a free parameter as well as $\mu_0$. This is because there are many possibilities to generate it whether relating to the $\mu$ problem or not. For example, we can consider that a sequestered sector has SUSY breaking of ${F_{\rm seq} \over M_{\rm pl}}\sim \mathcal{O}$(1)TeV and the supergravity potential automatically contains $B_0 \sim {F_{\rm seq} \over M_{\rm pl}}$. 
It is confirmed by FeynHiggs 2.11.2 \cite{Hahn:2013ria} that the spectrum in Tab.\ref{tab:IRspectrum} leads to the Higgs boson mass, $m_h =$125.7 GeV, and the SUSY contribution to the muon $g-2$, $\delta({g-2\over 2})=2.6 \times 10^{-9}$. The GCU is obtained at the SUSY GUT scale perturbatively, with $\alpha_{\rm GUT}^{-1} \sim 8-10$. Therefore, the model can explain the muon $g-2$ anomaly naturally within the $1~\sigma$ level and the Higgs boson mass with $\mathcal{O}(10)$ TeV stops. 
If the stop masses are heavier than 20 TeV, the calculation here is not reliable since the mediation of $\Lambda_d$ to the L sector via the H sector (Fig.\ref{fig:GMSB}) is not negligible compared with the mediation of $\Lambda$. Naively, this does not occur when
\begin{align}
\label{eq:consy}
m_{{\tilde e}_2}^2>{1\over 16\pi^2} m_{{\tilde e}_3}^2.
\end{align}
The condition also guarantees the neglecting of the other sub-dominant factors, such as the 2-loop RGE effects, 
the natural deviation of couplings from the $N$=2 SUSY relations (Sec.\ref{chap:mech}), and up to factor $\sim$3 differences between the SUSY masses $M$, $Z$ and $Z_d$ (Sec.\ref{chap:Model}).
This is the constraint for the right boundary of the allowed region in Fig.\ref{fig:GMSB}.

\begin{table}[!h]
\begin{center}
\begin{tabular}{|c|c|c|c|c|c|c|}\hline Parameter & $\tan\beta$&$M$&$\Lambda_d$&$\Lambda$& $\mu_0$&$B_0$\\\hline 
Value & 7 & $10^{10}$ GeV&930 TeV&68 TeV&12 TeV&1.5 TeV\\\hline \end{tabular}
 \caption{The fundamental parameters corresponding to the black dot in Fig.\ref{fig:GMSB}}
 \label{tab:Fpara}
\end{center}
\end{table}

\begin{table}[!h]
\begin{center}
\begin{tabular}{|c|c|}\hline H sector & Mass \\\hline $\tilde{Q}_3$ & 12.5 TeV \\\hline $\tilde{u}_3$ & 10.9 TeV \\\hline $\tilde{d}_1, \tilde{d}_2, \tilde{d}_3$ & 14.1 TeV \\\hline $\tilde{e}_1, \tilde{e}_3$ & 2.6 TeV \\\hline $\tilde{Q}_1,\tilde{Q}_2$ & 14.1 TeV \\\hline $\tilde{u}_1,\tilde{u}_2$ & 14.2 TeV \\\hline A-Higgs & 11.4 TeV \\\hline Higgsino &  11.3 TeV \\\hline Higgs (by FH \cite{Hahn:2013ria}) & 125.7 GeV \\\hline 
 \end{tabular}
\begin{tabular}{|c|c|}\hline L sector & Mass \\\hline $\tilde{L}_3$ & 538 GeV \\\hline $\tilde{L}_1,\tilde{L}_2$ & 544 GeV \\\hline $\tilde{e}_2$ & 216 GeV \\\hline 
Mediator & Mass \\\hline $\tilde{G}$ & 1940 GeV \\\hline $\tilde{W}$ & 543 GeV \\\hline $\tilde{B}$ & 165 GeV \\\hline bini & ${\mathcal{O}}(10)$ MeV\\\hline sb & 2600 GeV \\\hline 
\end{tabular} 
\caption{The low energy spectrum corresponding to the black dot in Fig.\ref{fig:GMSB}}
\label{tab:IRspectrum}
\end{center}
\end{table}

There are six fundamental parameters, ($M$, $\tan\beta$, $\Lambda_d$, $\Lambda$, $\mu_0$, $B_0$), and three equalities: the Higgs boson mass and the electroweak symmetry breakdown conditions. Hence, three free parameters remain while the number of the sparticles lighter than 2 TeV are seven. Therefore, the model is predictive. We note that the predictability holds as long as Eq.(\ref{eq:consy}) is satisfied roughly, otherwise not only higher loop effects but also many subdominant parameters, which we have neglected, become effective. 
In our fundamental parameter set in Tab.\ref{tab:Fpara}, we can show only the mass of $\tilde{e}_2$ may alter at a factor of $\lesssim \mathcal{O}(1)$ from our prediction due to the closeness to the bound, Eq.(\ref{eq:consy}), while the other masses change at most by $\mathcal{O}(1)\%$ from a naive estimation of higher loop effects. 
Another feature of the spectrum is the heavy 1st/3rd generation right-handed sleptons. The model is favored by the CP bounds of the 1st generation fermion and the vacuum stability of the stau.

The model predicts new singlet boson and fermion: sb and bini. The effective couplings between the sb and the MSSM superfields are proportional to ${1 \over M^2}$ at the SUSY limit due to the $Z_2$ parity. The suppressed couplings 
imply the difficulty to detect them in experiments and the ease to be free from experimental 
problems. Cosmologically, the sb decay may spoil the successful BBN prediction. 
To avoid this, there are two possibilities: the cases of a low enough reheating temperature (or a high enough messenger scale) and of a low enough messenger scale. The former one prevents thermal generation of the sb while the latter one advances the decay before the BBN. 

We have not considered the constraints from cosmology in choosing the spectrum. Naively, the spectrum is not viable cosmologically, at least the thermal abundance of the bino-like neutralino is over-produced. However, $R$-parity violations (in which case the bini can be the dark matter), coannihilations with sfermions taking the higher corrections into account, dilutions via decay of a scalar field (which can be the sb) etc., can cure the problem. The cosmological implications in detail will be studied in our future work.

\setcounter{footnote}{0}
\section{Conclusions}
In the paper, we consider the situation that the SUSY scale is far beyond the EW scale, while some of the sfermions are light enough to deviate the measured values of some 
flavor physics from the SM predictions. 

We suggest a natural split mechanism using $N$=2 SUSY.  We show that $N$=2 SUSY protects the K{\" a}hler kinetic term of a 
hypermultiplet from quantum corrections even when SUSY is slightly broken. Therefore, partly $N$=2 SUSY models realize the natural split spectrum of sfermions, if some of the sfermions are contained in the hypermultiplets. The situation may be feasible in some underlying theories. 

We assume at a scale partly $N$=2 SUSY is realized, and build a natural split GMSB model that explains both the muon $g-2$ anomaly at the 1 $\sigma$ level and the measured Higgs 
boson mass. The model contains three free parameters and seven testably light sparticles with the masses below 2 TeV. In addition there is a new light fermion, bini. The bini has the mass of MeVs, which is 2-loop suppressed compared with the bino mass. 
The bini is experimentally safe due to the smallness of the effective couplings.

Therefore, we have shown that a partly $N$=2 SUSY extension is a 
natural split mechanism for sfermions, and the light sfermion, if detected, is not only 
a sign of $N$=1 SUSY but also may be a sign of $N$=2 SUSY.

\section{Discussions}
Let us discuss about the assumptions we have made to build the natural split GMSB model. 
We have assumed the following conditions for the sake of clarity as a good
example of the natural split mechanism.

We have chosen the additional fields to the MSSM that can be put into complete SU(5) multiplets. However, 
there are other choices of fields whether preserving GCU at the SUSY GUT scale or not.

We have assumed that the low energy spectrum is the MSSM-like and is calculated by using the 
RGEs of the MSSM, which constrains the UV spectra into ones that are stable under the RGEs from 
the viewpoint of naturalness. However more various stable UV spectra can exist, especially if some of the $N
$=2 partners do not decouple, since a part of the ${\rm SU(2)_R}$ would cancel some of the RGE 
corrections. In the case, the experiments constrain the model more severely, but plenty of unique 
signatures should be possible to predict.

As explained in the Sec.\ref{chap:mech}, we should have loop suppressed deviations from the exact SO(2) relations in the L and H sectors from the viewpoint of naturalness. The sfermion mass corrections due to the deviations are the same order as the ones via the H sector in the loop expansion, hence if the expansion is valid, the effects are neglected due to Eq.(\ref{eq:consy}). 
 However, we have to face with the breakdown of the validity of loop expansion due to the hierarchy of couplings 
 and flavors. In our case, we can not put all the missing partners just above our messenger scale due to the 
 assumption of the GCU. If we put the missing partners at the SUSY GUT scale and assume the partly $N$=2 
 SUSY condition there, the $N$=2 relations would be broken due to the invalidity of the loop expansion after 
 ressumming the logarithms by using the RGEs of Eq.(\ref{eq:L}). This problem can be solved by setting the 
 messenger scale near the SUSY GUT scale in which case the stop masses are reduced to $\sim 7$TeV to solve 
 the muon $g-2$ anomaly within 1$\sigma$ level which is less favored in the light of the CP, flavor and proton  
 decay problems. Also perturbativity does not constrain the additional particle contents any more. 
 We can also assume some underlying theory above the messenger scale to modify the RGEs, however in the 
 case we have to answer why the theory does not change the predictions of the model. 
 Therefore, this is a naturalness problem above the messenger scale. Conversely, this is also a hint to constrain 
 the additional particles, since our model is not the only viable one. 

 In the paper, from the viewpoint of naturalness at and below the messenger scale, we have built one of the 
 viable models. However, the underlying theory and the naturalness above the messenger scale should further 
 constrain the additional particle contents. And, improvements will be shown in our future works as well as their phenomenology and cosmology in detail.

\section*{Acknowledgement}
W.Y. would like to thank  Masahiro Yamaguchi for the continuing guidance and the encouragement during his Ph.D. This research is supported in part by the ministry of education, culture, sports, science and technology of Japan.

\providecommand{\href}[2]{#2}\begingroup\raggedright\endgroup

\end{document}